# Assessing the learning behavioral intention of commuters in mobility practices

Waqas Ahmed [A,1], Habiba Akter [2], Sheikh M. Hizam [3], Ilham Sentosa [4] and Syeliya Md. Zaini [5]

*UniKL Business School (UBIS), Universiti Kuala Lumpur, Kuala Lumpur, 50300, Malaysia.*

**Abstract**
Learning behavior mechanism is widely anticipated in managed settings through the formal syllabus. However, heading for learning stimulus whilst daily mobility practices through urban transit is the novel feature in learning sciences. Theory of planned behavior (TPB), technology acceptance model (TAM), and service quality of transit are conceptualized to assess the learning behavioral intention (LBI) of commuters in Greater Kuala Lumpur. An online survey was conducted to understand the LBI of 117 travelers who use the technology to engage in the informal learning process during daily commuting. The results explored that all the model variables i.e., perceived ease of use, perceived usefulness, service quality, and subjective norms are significant predictors of LBI. The perceived usefulness of learning during traveling and transit service quality has a vibrant impact on LBI. The research will support the informal learning mechanism from commuters' point of view. The study is a novel contribution to transport and learning literature that will open the new prospect of research in urban mobility and its connotation with personal learning and development.

**Keywords 1**
Learning Behavioral Intention, Mobility, Service Quality.

## 1. Introduction

On the internet, the word "nostalgia" is triggered towards the remembrance of old-ideal-days by watching certain photos, animation, or movies, etc. however this word was merely considered for homesickness till last century. The rapid transformation in living matters of life is the main cause of such gut feelings. Similarly, these changes necessitate a prompt response from humans towards their professional and personal life also. Learning through schoolings and graduating with a degree enables one to survive in professional and social life. However, the pace technology and innovation have adopted and the way day-to-day novel methods and processes of modernization are making obsolete the older ones, merely formal education cannot guarantee a prosperous future for the individuals. It reckons as go with the flow analogy. The learning process needs to be honed by the individual for creating the breakeven point between new technology integration and his/her skills development. To do so, informal learning could facilitate by enabling the skills learning process in personal and common settings. Due to the acceptance of technology for learning, portable digital devices are controlling the e-learning process. The learners can use such electronic devices at any place and situation to make things happen. Informal learning can be a great deal towards personalized learning such as language learning, software learning for education or profession, music learning, or any type of skill learning and polishing. The means by which informal learning is feasible are multifaceted such as watching learning videos on YouTube, engaging the MOOC activities, attending the online learning session, listening to podcasts, etc. All such happenings are possible in outdoor formats during leisure time such as resting in the gardens, regular jogging in parks, and daily traveling by bus and train. In the case of urban localities, the daily commuting takes ample time to reach the destination for university or office personnel due to congestion and lack of personal conveyance.

Urban planning departments strive to provide sustainable transport facilities to citizens around the globe. The purpose is to enhance the quality of life by eradicating mobility poverty due to the higher urbanization levels. It includes the safe and economical transport system such as metro, light rail transit (LRTs), intercity bus/train transit services, etc. The quality of such services is based on convenience, accessibility, easiness, etc. As daily commuters receive convenient and safe amenities during their daily traveling, the leisure mechanism is developed in this process that signals towards engagement in the favorite hobby or learning of a certain purpose by the informal way [1]. By way of the daily commuters' time is occupied during transit services, to understand their behavior pattern would help to understand and improve travel behavior. Towards learning, the use of the mobile, laptop, and other gadgets have made the process convenient, however, there is a paucity of research about commuters' engagement of leisure time during traveling towards learning mechanism. This study endeavors to answer the expression as, do the travelers involve in informal learning during traveling? and what are their behavioral aspects that support this progression?.

To understand such behavior of learning during mobility practices, this study presents the mechanism of behavioral modeling. The theory of planned behavior and technology acceptance model are the most considerate theories to predict the behavior of users towards informal learning [2][3] and traveling behavior [4] [5]. Moreover, the quality of services rendered [6] also impacts the behavior of commuters during travel [7]. However, to understand the scenario of mobility practices and learning behavior, the evidence of such related research is lacking in the previous literature. Therefore, this study comprehends the behavioral assessment through TAM and TPB with perceived usefulness (PU), perceived ease of use (PE), subjective norms (SN), learning behavioral intention (LBI), and service quality(SQ) features of transit services. The research will bring an understanding of commuters' informal behavior towards personal learning and development. It will support research on transport service usability and human behavior.

## 2. Literature Review

Predicting and determining users' behavior and users' acceptance of services have become pivotal points for both academics and practitioners in recent times. Prior researchers used various theories to understand user behavior and their perceptions of the system. Among these, the technology acceptance model (TAM) has been broadly applied to elucidate the individual's intention using specific services in numerous areas, including transportation systems [8]. Given the successful outcomes of TAM at assessing user behavior and their perceptions of services and systems, we have chosen the TAM as an extensive model to illuminate transport users' behavior towards their learning intention while traveling. Besides, informal learning by technology is a novel topic, and hence it is suitable to be assessed by adopting the TAM model.

Davis originally proposed the TAM model, while later, Davis and his colleagues presented the TAM as a more advanced model [9], [10]. In general, TAM is used as a robustly validated model to predict and measure human behavior and their technology adoption process. TAM includes two particular constructs, including perceived ease of use (PE) and perceived usefulness (PU), and confirms that user behavior is predicted by PE and PU of the system usage [9], [10]. In TAM, PE is referred to as the individual perception to use any information technology that will be effort-free [9]. Based on this definition, this study signifies PE as the transport users' perception of informal learning that the mode of transport should be easy to use. On the other hand, PU is meant as the users' perception to use any information system that will boost or foster their performance level [9]. Applying this concept, this study defines PU as an individual's perception of informal learning which will improve his/her learning performance during the time spent to travel. The strengthening of these perceptions leads to a positive behavioral intention of informal learning, thereby PE and PU are considered as strong predictors of travelers' LBI. Using an online survey among 1045 respondents, a group of scholars found that PU had positive effects on user behavioral intention [8]. Within the U.S. context, another research analyzed the collected data including 377 responses, in which the results showed that both PE and PU positively impacted user behavior using any system [11]. Furthermore, earlier researchers empirically proved that PE and PU were positively and significantly associated with individual attitude and behavioral intention [12]. Therefore, the following hypotheses have been developed:

**Hypothesis 1 (H1):** PU has a positive connection with commuters' LBI.

**Hypothesis 2 (H2):** PE is positively associated with commuters' LBI.

In the last two decades, many studies have been conducted on service quality (SQ) towards behavioral intention in the public transport context [7], [13], [14]. The term "SQ" is the outcome of consumers' evaluation, which differentiates between their expectations for a service offering and their perceptions of specific service receiving [15]. Overall, the notion of SQ is crucial to effective business outcomes because qualified service has a fruitful impact on customers' decision-making. According to [15], SQ is a multifaceted phenomenon encompassing three unique features, namely intangibility, heterogeneity, and inseparability. These characteristics of services in the context of travelers' perceptions, such as intangibility (safe and comfortable journey); heterogeneity (frequency or number of daily services, speed, punctuality, good connection with another mode of transport, and cost); and inseparability (repeatedly service delivered, service providers' awareness to the consumers) lead to the measurement of individual behavior [7]. However, the findings of existing studies found a positive and significant association between SQ and consumers' behavioral intentions. For example, surveying public transport users in five European cities (i.e., Madrid, Rome, Berlin, Lisbon, and London), recent research indicated that SQ delivered by a transport provider was positively associated with users' behavioral intention [7]. Similarly, the link between SQ and transport users' behavioral intention was found to be significant in another research [14]. Hence, when transport providers seem to offer tranquil and enjoyable services to commuters, they will be more keenly inclined to learn through their digital learning materials. Based on this, our next hypothesis is:

**Hypothesis 3 (H3):** SQ positively influences commuters' LBI.

TPB was initiated firstly by Ajzen in the early 1990s [16] and now it has been extensively applied as a theory on human behavior due to its robust control predicting individuals' behavioral intention [5], [17], [18]. According to TPB, human behavioral intention is determined by three influential factors, namely attitude, subjective norm, and perceived behavior control. In this study, the subjective norm (SN) is taken into consideration assessing learning behavior intentions (LBI) among travelers in the mode of public transport. Within TPB, the second predictive factor for behavioral intention is the SN. The SN is defined as "the individual perception towards social pressure either performing or not performing the behavior" [16]. In other words, the SN coincides with a person's belief to comply whether the social reaction would appropriate or deter his/her behavior [18]. The higher subjective norms are perceived by individuals, the more likely to behave in a certain manner [5], [19]. With reference to the survey on 1333 respondents, a recent study's findings revealed that SN positively predicted behavioral intention [17]. In our research, the SN is referred to as the commuters tend for LBI in public transport. If a person understands that a group of people support his/her behavior while on public transport, he/she may be more pursued and intended to learn something relying on social media or any learning website. Thus, it is postulated the following hypothesis:

**Hypothesis 4 (H4):** SN positively predicts commuters' LBI in public transportation.

The conceptual framework according to the given literature is illustrated in Figure1.

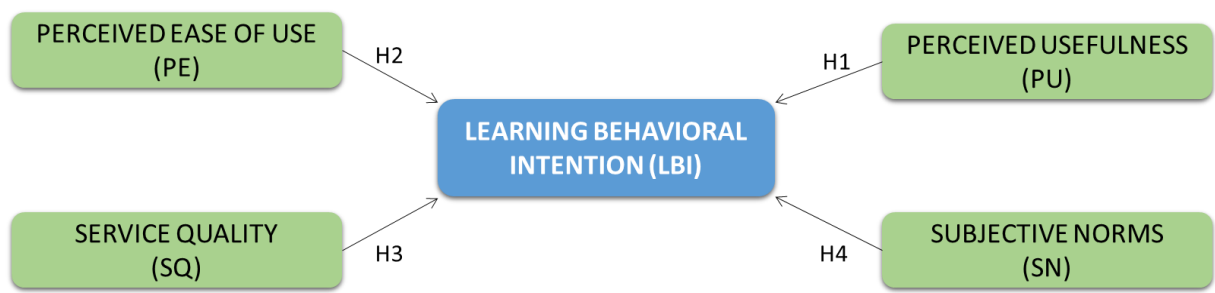

**Figure 1. Conceptual Framework**

## 3. Methodology

Behavioral assessment refers to the technique and methods relating to causation and according to Saunders' research onion [20], the positivist philosophy with the deductive theoretical approach is suitable in cause and effect analysis through the quantitative method. This research being the causal assessment of behavior had adopted the positivist philosophy with the deductive approach by considering the TAM and TPB as the theoretical background. To steer the analysis, a cross-sectional survey to collect the data was considered. Data was collected through a close-ended questionnaire. The questionnaire consisted of two sections, comprising of demographic information section and variable items section. The questionnaire items were adopted from previous studies as for service quality 14 items were taken from a public transport research[7], perceived usefulness (PU), perceived ease of use (PE), and learning behavioral intention (LBI) were taken from [9] and subjective norms (SN) was adopted from [21]. To gather the response of respondents for questionnaire items, a 5-point Likert scale was used with "strongly disagree = 1" to "strongly agree = 5". For this study, the snowball sampling method was used through online questionnaires because the current pandemic circumstances hindered the free movement and contact with the respondents physically on the transit stations. The research was conducted in Malaysia's largest metropolitan vicinity, Klang Valley or Greater Kuala Lumpur (i.e., Kuala Lumpur and its surrounding districts), which is equipped with an advanced rail transit network that provides sustainable mobility facilities [22]. Prior to answering the questionnaire, respondents were asked for the basic requirement that either in common days they travel daily basis through public transit system i.e., KTM, LRT, MRT, Monorail, in the Klang Valley or not. Upon positive answer, the were allowed to provide their responses. Data of 117 travelers were collected during the 4 weeks survey. The sample size was enough as per the number of latent carriable to run the basic behavioral assessment analysis. The data was then sorted for demographic results, reliability and validity, factor loadings, and model fit analysis. Data was then analyzed the Structural Equation Modeling (SEM). Analysis process was carried out by SPSS Statistics v25 and SPSS AMOS v23.

## 4. Results

Post data collection, initial steps include analyzing the data for demographic classification and it was explored that the majority of travelers of recorded data as per demographic variables consisted of male 69%, "26-35 years" age group 38%, education qualification 52%, travelers employees 59%, and 40% respondents spend around 2 hours in daily commuting. The detailed demographic numbers are illustrated in Table 1. SEM is based on two models named as the measurement model to ensure the data accuracy within items of a variable and structural model to assess the regression path among the model variables. To process the data for behavioral analysis, according to the measurement model of SEM, reliability, and validity of collected data were assessed and all the variables found reliable and valid at the threshold of "greater than 0.70". In Table 2, reliability by Cronbach's alpha > 0.70 and Validity with Composite reliability (CR) > 0.70 and Average Variance Extracted (AVE) > 0.40 showed the data valid for causal analysis[23]. Moreover, the outer loadings of each item was also calculated to comprehend the instrument's correlation towards its respective variable with a value higher than 0.50, and according to Table 2, all the outer loadings of items met the minimum threshold of 0.50. To realize the correlation impact between the model variables, the correlation matrix, as presented in Table 3, proved that model variables had the minimum value of correlation among each other i.e., less than 0.90. Prior to the structural analysis of the model, according to [23] the model good fit indices were tested and all the indices found fit to proceed for path analysis as shown in Table 4. The interpretation of results is based on the recommended parameters. In the structural model of SEM, the hypothesis testing was conducted by path analysis. The results of all four hypotheses found significant with critical ratio (CR) or t-statistics value greater than 1.96 (i.e., CR>1.96) and significance level is less than 0.05 (i.e., P-value <0.05) [23], [24] while beta estimates of perceived usefulness (PU) appeared higher than other variables. The path analysis results are presented in Table 5 and the structural model is portrayed in Figure 2.

**Table 1**
Demographic Results

| Demographics | Specification (numbers) |
|---|---|
| **Gender** | Male = 81, Female = 36 |
| **Age** | "15-25 Years"=14, "26-35 Years"=45, "36-45 Years"=39, "More than 45"= 21 |
| **Education** | Diploma = 23, Bachelors = 61, Masters = 28, Doctorate and Higher = 05 |
| **Profession** | Students = 37, Employees = 69, Entrepreneur/Business Owner =11 |
| **Daily Transit** | 01 hour = 31, 02 hours = 47, More than 2 hours = 39 |

**Table 2**
Outer Loadings, Reliability, and Validity Results

| Variables | Items | Outer Loadings | Cronbach's Alpha | C.R | AVE |
|---|---|---|---|---|---|
| **Service Quality (SQ)** | SQ1 | 0.676 | 0.907 | 0.921 | 0.455 |
| | SQ2 | 0.746 | | | |
| | SQ3 | 0.672 | | | |
| | SQ4 | 0.669 | | | |
| | SQ5 | 0.628 | | | |
| | SQ6 | 0.710 | | | |
| | SQ7 | 0.701 | | | |
| | SQ8 | 0.703 | | | |
| | SQ9 | 0.702 | | | |
| | SQ10 | 0.577 | | | |
| | SQ11 | 0.584 | | | |
| | SQ12 | 0.690 | | | |
| | SQ13 | 0.692 | | | |
| | SQ14 | 0.669 | | | |
| **Perceived Ease of Use (PE)** | PE1 | 0.840 | 0.900 | 0.924 | 0.669 |
| | PE2 | 0.706 | | | |
| | PE3 | 0.867 | | | |
| | PE4 | 0.836 | | | |
| | PE5 | 0.845 | | | |
| | PE6 | 0.804 | | | |
| **Perceived Usefulness (PU)** | PU1 | 0.851 | 0.921 | 0.941 | 0.761 |
| | PU2 | 0.865 | | | |
| | PU3 | 0.911 | | | |
| | PU4 | 0.856 | | | |
| | PU5 | 0.876 | | | |
| **Subjective Norms (SN)** | SN1 | 0.901 | 0.922 | 0.944 | 0.810 |
| | SN2 | 0.945 | | | |
| | SN3 | 0.930 | | | |
| | SN4 | 0.818 | | | |
| **Learning Behavioral Intention (LBI)** | LBI1 | 0.926 | 0.954 | 0.967 | 0.879 |
| | LBI2 | 0.947 | | | |
| | LBI3 | 0.926 | | | |
| | LBI4 | 0.952 | | | |

**Table 3**
Correlation Matrix

|     | BI       | PE       | PU       | SN      | SQ |
|-----|----------|----------|----------|---------|----|
| BI  | 1        |          |          |         |    |
| PE  | 0.630388 | 1        |          |         |    |
| PU  | 0.66001  | 0.516962 | 1        |         |    |
| SN  | 0.477406 | 0.394593 | 0.312571 | 1       |    |
| SQ  | 0.544347 | 0.446157 | 0.524747 | 0.27873 | 1  |

**Table 4**
Model Fit Indices

| Measure | Estimate | Threshold        | Interpretation |
|---------|----------|------------------|----------------|
| CMIN    | 556.588  | --               | --             |
| DF      | 424      | --               | --             |
| CMIN/DF | 1.313    | Between 1 and 3  | Excellent      |
| CFI     | 0.946    | >0.95            | Acceptable     |
| SRMR    | 0.059    | <0.08            | Excellent      |
| RMSEA   | 0.052    | <0.06            | Excellent      |
| PClose  | 0.389    | >0.05            | Excellent      |

**Table 5**
Hypothesis Testing

|    | Hypotheses   | Estimate | S.E.  | C.R.  | P     | Result   |
|----|--------------|----------|-------|-------|-------|----------|
| H1 | PU → LBI     | 0.389    | 0.094 | 4.151 | 0.000 | ACCEPTED |
| H2 | PE → LBI     | 0.290    | 0.09  | 3.322 | 0.000 | ACCEPTED |
| H3 | SQ → LBI     | 0.274    | 0.12  | 2.289 | 0.022 | ACCEPTED |
| H4 | SN → LBI     | 0.212    | 0.076 | 2.802 | 0.005 | ACCEPTED |

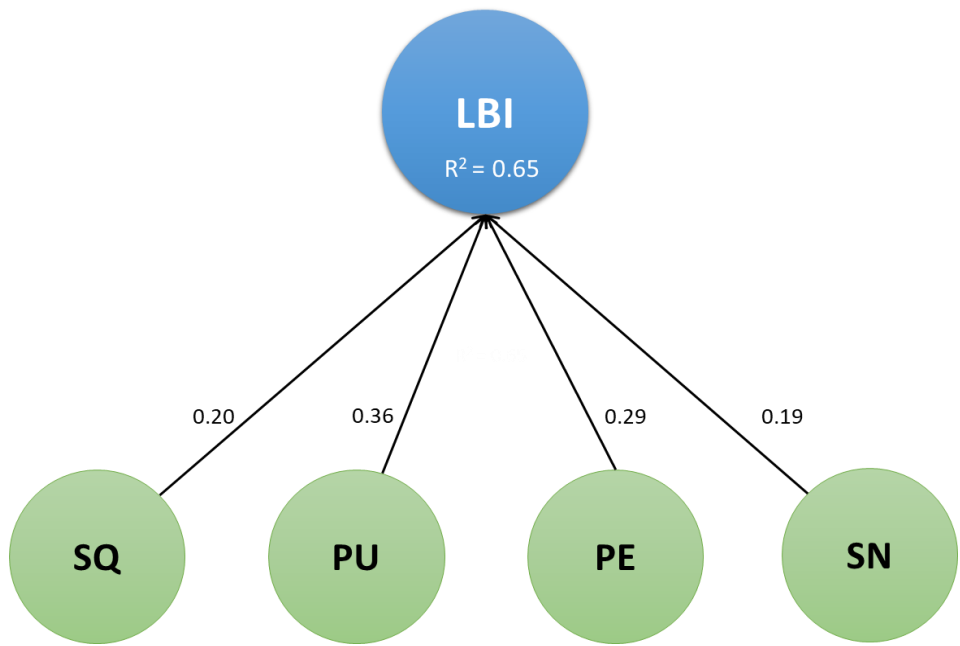

**Figure 2. Structural Model**

## 5. Conclusion

The purpose of this study was to elucidate the learning behavioral intention of travelers during their daily commuting through urban transit. In the last section, SEM analysis explored that all the variables such as service quality of public transport (SQ), subjective norms (SN), perceived usefulness (PU), and perceived ease of use (PE) are significant towards learning behavioral intention (LBI) during mobility practices. The overall model exhibited a change of 65% (i.e., R-squared = 0.65) which showed that all four variables possess quite an impact to change the behavior of commuters. Perceived usefulness was found as the most impactful factor towards learning behavioral intention (i.e., H1) with the beta estimate of 38.9%. It shows that commuters' perception of advantages by using technology for informal learning during daily mobility is quite robust. Similarly, when commuters find it easier to travel during daily commuting, as hypothesized by H2, the tendency to engage in the learning process would be more augmented. The important feature of service quality of transit services that explains the rendered services by travelers on daily basis and being the recipient of such services, do such services impact on behavioral intention towards informal learning. The hypothesis H3 of service quality towards BLI is significant and it had a reasonable impact on BLI with a beta estimate of 27.4%. The services features of daily used transit included operational hours, frequency of trains in normal and peak times, the proximity of transit areas, regularity, speed, fares rate, hygiene factors in the train, temperature-centric, safety features, individuality in the train, accessibility of services inside the train, security aspects, information and intermodally [7]. According to hypothesis 4, the impact of the views of family, peers, and friends to engage in learning during commuting has also been found significant. It showed positive social support and recommendation towards engaging in informal learning in traveling mode.

The inferences of research are the novel contribution to the literature and practice. The assessment of learning behavioral intention in commuting was the unexplored phenomenon as being the utmost capacity of leisure during the safe and sustainable daily traveling needs to be managed for creativity and development of individuals and professionals. The study explicitly demonstrated the trend of informal learning and development among the employees commuter that shows the encouraging perception of inclination towards achieving the breakeven point of skill development and work-related innovation pace [25], [26]. This learning process in mobility can be based on any form of learning activity by using mobile phone apps, even without the internet also, and the matter of significance of such behavior is to enable the instructors to frame the coaching scenarios for students in daily life. It can also support the individuals to enlist their projected skills to learn through their daily commuting. The conclusion of our research on this topic is the first step to understand the informal learning in travel behavior, the proceeding work can enhance the body of knowledge.

Though this research carried out vital insights into the certain key constructs which positively influence user perception of learning behavior, several limitations could be calibrated in any further study. First, we developed the research framework elucidating four determinants (i.e., PU, PE, SN, and SQ) of commuters' LBI. From the academic viewpoint, the proposed framework could be more elevated and used to assess user beliefs of informal learning opportunities in the travel mode. Both TAM and TPB may be further extended as a core notion integrating with the unified theory of acceptance and use of technology (UTAUT). Future studies may enlist the facilitating condition (FC) from UTAUT theory for understanding the commuters' state of traveling and surroundings towards behavior change[27]. Similarly adding the demographic features of commuters could bring more insight into LBI predictors. Second, as the studied sample consisted of 117 responses which are limited for the detailed SEM study, a well-structured quantitative survey with a larger sample size (e.g., >400) will contribute to more in-depth assessment through SEM analysis. Lastly, another limitation of the current research is linked to the web-based survey strategy using a snowball sampling technique, which may be inappropriate for providing a seamless representative of the respondents. Hence, more research is needed using the representative sampling technique i.e., systematic random sampling, stratified random sampling in the face-to-face survey.

## References.